\documentclass[12pt,english]{article}

\usepackage[T1]{fontenc} 
\usepackage[utf8]{inputenc}
\usepackage{geometry}
\usepackage{ulem}
\usepackage{amsmath}
\usepackage{graphicx}
\usepackage{setspace}
\usepackage{esint}
\usepackage{indentfirst}
\makeatletter

\providecommand{\tabularnewline}{\\}

\newcommand{\lyxaddress}[1]{
	\par {\raggedright #1
	\vspace{1.4em}
	\noindent\par}
}

\def\be{\begin{equation}}
\def\ee{\end{equation}}

\def\bea{\begin{eqnarray}}
\def\eea{\end{eqnarray}}

\def\ba{\begin{array}}
\def\ea{\end{array}}

\def\bc{\begin{center}}
\def\ec{\end{center}}

\def\bl{\begin{flushleft}}
\def\el{\end{flushleft}}

\def\br{\begin{flushright}}
\def\er{\end{flushright}}

\def\bi{\begin{itemize}}
\def\ei{\end{itemize}}

\def\bt{\begin{tabular}}
\def\et{\end{tabular}}

\numberwithin{equation}{section}

\usepackage{slashed}
\usepackage{hyperref}
\usepackage{graphicx}
\usepackage[numbers,sort&compress]{natbib}
\date{}

\hypersetup{colorlinks=true,citecolor=blue,linkcolor=red}

\makeatother

\usepackage{babel}
\begin{document}
\title{Instability of the regularized 4D charged Einstein-Gauss-Bonnet de-Sitter
black hole}
\author{Peng Liu$^\ast$, Chao Niu\footnote{
  phylp@jnu.edu.cn,\,\, niuchaophy@gmail.com}, Cheng-Yong Zhang\footnote{
  corresponding author, zhangcy@email.jnu.edu.cn}
%\thanks{phylp@jnu.edu.cn, niuchaophy@gmail.com, }
}

\maketitle

\lyxaddress{\begin{center}
\textit{Department of Physics and Siyuan Laboratory, Jinan University,
Guangzhou 510632, China}
\par\end{center}}
\begin{abstract}
We studied the instability of the regularized  4D charged Einstein-Gauss-Bonnet
de-Sitter black holes under charged scalar perturbations. The unstable
modes satisfy the superradiant condition, but not all modes satisfying
the superradiant condition are unstable. The instability occurs when
the cosmological constant is small and the black hole charge is not
too large. The Gauss-Bonnet coupling constant makes the unstable black
hole more unstable when both the black hole charge and cosmological
constant are small, and makes the stable black hole more stable when
the black hole charge is large.
\end{abstract}

\section{Introduction}

It is known that general relativity should be modified from both the
viewpoint of theory and observation. For examples the general relativity
can not be renormalized and can not explain the dark side of the Universe.
On the other hand, the Lovelock theorem states that in four dimensional
vacuum spacetime, the general relativity with a cosmological constant
is the unique metric theory of gravity with second order equations
of motion and covariant divergence-free \cite{Lovelock1971}. Beyond
general relativity, one must go to higher dimensional spacetime, or
add extra fields, or allow higher order derivative of metric or even
abandon the Riemannian geometry. Various modified gravity theories
have been proposed \cite{Clifton2011}. In this work, we focus on
the Einstein-Gauss-Bonnet (EGB) gravity. The EGB gravity is one of
most promising candidates of modified gravity. It has second order
equations of motion and free of Ostrogradsky instability \cite{Woodard2015}.
It appears naturally in the low energy effective action of heterotic
string theory \cite{Gross1987}. However, the Gauss-Bonnet (GB) term
has nontrivial dynamics only in higher dimensions in general. In four
dimensional spacetime, it is a topological term and does not contribute
to the dynamics. By rescaling the GB coupling constant in a special
way, novel four dimensional GB black hole solution was found recently
\cite{Glavan20192019a}. This work provides a new classical four dimensional
gravity theory and has inspired many studies, including the new black
hole solutions \cite{Fernandes2003,4DEGBSolutions,Kobayashi2004},
perturbations \cite{4DStability,Zhang2020Hawking,Zhang2020Super,Cuyubamba2020},
shadow and geodesics \cite{4DEGBShadow}, thermodynamics \cite{4DEGBThermo},
and other aspects \cite{4DEGBCollapse,4DEGBOthers}.

The novel black hole solution has some remarkable properties. Its
singularity at the center is timelike. The gravitational force near
the center is repulsive and the free infalling particles can not reach
the singularity \cite{Glavan20192019a}. One may expect that the novel
black hole solution would also show some new properties under perturbations,
and some related works have been done \cite{4DStability,Zhang2020Super}.
The study of the stability of black hole is an active area in black
hole physics. It can be used to extract the parameters of the black
hole such as its mass, charge and angular momentum. The stability
of black hole is also related to gravitational wave, black hole thermodynamics,
information paradox and holography, etc. \cite{Konoplya2011}. Among
these studies, the stability of black hole in asymptotic de Sitter
(dS) spacetime is intriguing. The Kerr black hole in dS spacetime
also behaves very different with those in asymptotic spacetime under
perturbations \cite{Zhang2014}. The four dimensional Reissner-Nordstr\"om-de
Sitter (RN-dS) may violate the strong cosmic censorship \cite{Cardoso2017,Zhong2019}.
The higher dimensional RN-dS and Gauss-Bonnet-de Sitter (GB-dS) black
holes are unstable \cite{Konoplya2008,Li2019RNdS}.

A quite surprising and still not very well understood result was discovered
in \cite{Zhu2014,Konoplya2014b,Destounis2019}, where they showed
that RN-dS black hole is unstable under charged scalar perturbations.
Such instability satisfies superradiance condition \cite{Brito2015}.
However, only the monopole $l=0$ suffers from this instability. Higher
multipoles are stable. This is distinct from superradiance. To understand
the precise mechanism, one may need the nonlinear studies, which was
partially answered in recent works \cite{Zhong2019}. In this paper,
we consider the instability of the novel 4D charged EGB black hole
in asymptotic dS spacetime under charged scalar perturbations. We
will see that the behavior is very different with the case in asymptotic
flat spacetime which has been done in \cite{Zhang2020Super}. The
Gauss-Bonnet coupling constant plays a more subtle role here.

The paper is organized as follows. Section 2 describes the novel 4D
EGB-RN-dS black hole and gives the reasonable parameters region. Section
3 shows the charged scalar perturbation equations. Section 4 describes
the numerical method we used and gives the results of the quasinormal
modes. Section 5 is the summary and discussion.

\section{The 4D charged EGB-dS black hole }

The action of the EGB gravity with electromagnetic field in $D$-dimensional
spacetime has the form
\begin{equation}
S=\frac{1}{16\pi}\int d^{D}x\sqrt{-g}\left[R+2\Lambda+\frac{\alpha}{D-4}\mathcal{G}^{2}-F_{\mu\nu}F^{\mu\nu}\right],\label{eq:action}
\end{equation}
 with $R$ the Ricci scalar, $g$ the determinant of the metric $g_{\mu\nu}$,
and $\Lambda$ the positive cosmological constant. The Maxwell tensor
$F_{\mu\nu}=\partial_{\mu}A_{\nu}-\partial_{\nu}A_{\mu}$, in which
$A_{\mu}$ is the gauge potential. The Gauss-Bonnet term
\begin{equation}
\mathcal{G}^{2}=R^{2}-4R_{\mu\nu}R^{\mu\nu}+R_{\mu\nu\alpha\beta}R^{\mu\nu\alpha\beta}=\frac{1}{4}\delta_{\rho\sigma\gamma\delta}^{\mu\nu\alpha\beta}R_{\ \ \mu\nu}^{\rho\sigma}R_{\ \ \alpha\beta}^{\gamma\delta},
\end{equation}
 with $R_{\mu\nu}$ the Ricci tensor and $R_{\mu\nu\alpha\beta}$
the Riemann tensor. We have rescaled the Gauss-Bonnet coupling constant
$\alpha$ by a factor $\frac{1}{D-4}$ in (\ref{eq:action}).

Varying the action with respect to the metric, one gets the equation
of motion
\begin{equation}
G_{\mu\nu}+\frac{\alpha}{D-4}H_{\mu\nu}=T_{\mu\nu}+\Lambda g_{\mu\nu}.\label{eq:EinsteinEq}
\end{equation}
Here $G_{\mu\nu}$ is the Einstein tensor and the contribution from
the GB term is given by
\begin{equation}
H_{\mu\nu}=2(RR_{\mu\nu}-2R_{\mu\sigma}R_{\ \nu}^{\sigma}-2R_{\mu\sigma\nu\rho}R^{\sigma\rho}-R_{\mu\sigma\rho\beta}R_{\ \ \ \nu}^{\sigma\rho\beta})-\frac{1}{2}g_{\mu\nu}\mathcal{G}^{2}.
\end{equation}
 In general, $H_{\mu\nu}$ vanishes in four dimensional spacetime
and does not contribute to the dynamics. However, it was argued that
the vanishing of $H_{\mu\nu}$ in four dimension might be canceled
by the rescaled GB coupling constant $\frac{\alpha}{D-4}$, and new
solutions were found \cite{Glavan20192019a}. The energy-momentum tensor
of the Maxwell field in (\ref{eq:EinsteinEq}) takes the form
\begin{equation}
T_{\mu\nu}=\frac{1}{4}\left(F_{\mu\sigma}F_{\ \nu}^{\sigma}-\frac{1}{4}g_{\mu\nu}F_{\alpha\beta}F^{\alpha\beta}\right).
\end{equation}
 In spherical symmetric spacetime, the electrovacuum solution of (\ref{eq:EinsteinEq})
can be written as

\begin{equation}
ds^{2}=-f(r)dt^{2}+\frac{1}{f(r)}dr^{2}+r^{2}(d\theta^{2}+\sin^{2}\theta d\phi^{2}),\label{eq:metric}
\end{equation}
 where the metric function
\begin{equation}
f(r)=1+\frac{r^{2}}{2\alpha}\left(1-\sqrt{1+4\alpha\left(\frac{M}{r^{3}}-\frac{Q^{2}}{r^{4}}+\frac{\Lambda}{3}\right)}\right),
\end{equation}
 and the gauge potential
\begin{equation}
A=-\frac{Q}{r}dt.
\end{equation}
 Here $M$ is the black hole mass and $Q$ the black hole charge.
When $\alpha\to0$, this solution returns to the RN-dS black hole.
As $r\to\infty$, it gives the asymptotic dS spacetime with an effective
positive cosmological constant. Note that the solution (\ref{eq:metric})
coincides formally with the ones obtained from conformal anomaly and
quantum corrections \cite{Cai20019} and those from Horndeski theory
\cite{Kobayashi2004}.

The parameters $M$ and $Q$ are positive. The GB coupling constant
$\alpha$ can be either positive or negative here. In appropriate parameter
region, the solution has three horizons: the inner horizon $r_{-}$,
the event horizon $r_{+}$ and cosmological horizon $r_{c}$. For
negative $\alpha$, the metric function may not be real in small $r$
region. But since we are only interested in the region between $r_{+}$
and $r_{c}$, we allow negative $\alpha$ in this work. For convenience,
hereafter we fix the black hole event horizon $r_{+}=1$. Then the
mass parameter can be expressed as
\begin{equation}
M=1-\frac{\Lambda}{3}+Q^{2}+\alpha.
\end{equation}
\begin{figure}[htbp]
	\centering
	\includegraphics[scale=0.5]{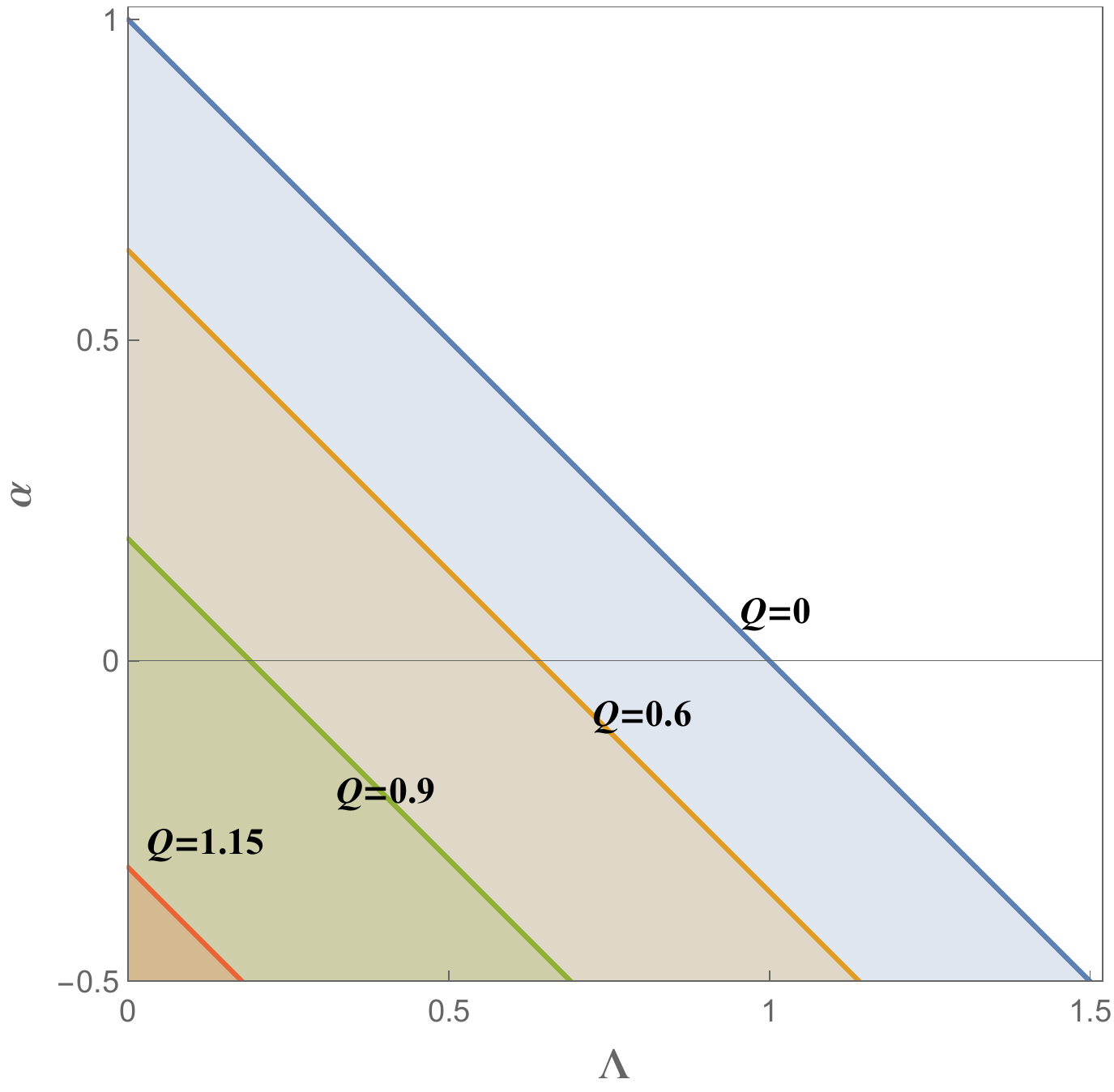}
	\caption{The parameter region that allows the event horizon $r_{+}$ and cosmological horizon $r_{c}$. The region is bounded by $Q^{2}+\alpha+\Lambda<1,$ $-0.5<\alpha$, $\Lambda>0$ and Q>0. As $Q$ increases, the allowed region for $(\Lambda,\alpha)$ shrinks. The extremal value of $Q=\sqrt{3/2}$.}
	\label{fig:ParameterRegion}
\end{figure}
Note that to ensure $f(1)=0$, there must be $\alpha>-1/2$. The parameter
region where allows the black hole event horizon $r_{+}$ and the
cosmological horizon $r_{c}$ can be determined by requiring $f'(r_{+})>0$,
which implies the black hole temperature is positive. This leads to
\begin{equation}
Q^{2}+\alpha+\Lambda<1.
\end{equation}
 This formula is very similar to the neutral case \cite{Zhang2020Hawking}.
The final parameter region is shown in Fig. \ref{fig:ParameterRegion}.

\section{The charged scalar perturbation}

It is known that fluctuations of order $\mathcal{O}(\epsilon)$ in
the scalar field in a given background induce changes in the spacetime
geometry of order $\mathcal{O}(\epsilon^{2})$ \cite{Brito2015}.
To leading order we can study the perturbations on a fixed background
geometry. Let us consider a massless charged scalar field $\psi$
on the background (\ref{eq:metric}). Its equation of motion is
\begin{equation}
0=  D^{\mu}D_{\mu}\psi\equiv g^{\mu\nu}\left(\nabla_{\mu}-iqA_{\mu}\right)\left(\nabla_{\nu}-iqA_{\nu}\right)\psi,\label{eq:perturEq}
\end{equation}
 where $q$ is the scalar charge and $\nabla_{\mu}$ the covariant
derivative. For generic background, we can take the following
decomposition
\begin{equation}
\psi=\sum_{lm}\int d\omega e^{-i\omega t}\frac{\Psi(r)}{r}Y_{lm}(\theta,\phi).
\end{equation}
 Here $Y_{lm}(\theta,\phi)$ is the spherical harmonics on the two
sphere $S^{2}.$ The angular part and the radial part of the perturbation
equation (\ref{eq:perturEq}) decouple. What we are interested in
is the radial part which can be written as the Schrödinger-like form
\begin{equation}
0=\frac{\partial^{2}\Psi}{\partial r_{\ast}^{2}}+\left(\omega^{2}-\frac{2qQ}{r}\omega-V_{\text{eff}}\right)\Psi,\label{eq:radEq}
\end{equation}
where the tortoise coordinate $dr_{\ast}=dr/f$ is introduced. The
effective potential reads
\begin{equation}
V_{\text{eff}}=-\frac{q^{2}Q^{2}}{r^{2}}+f\left(\frac{l(l+1)}{r^{2}}+\frac{\partial_{r}f}{r}\right).
\end{equation}
Unlike to the case in asymptotic flat spacetime where only one potential
barrier appears between $r_{+}$ and $r_{c}$, here a negative effective
potential well may appear between $r_{+}$ and $r_{c}$. We will see
that this potential well plays an important role in the instability
of charged EGB-dS black hole under perturbations.

The radial equation has following asymptotic behavior near the horizons.
\begin{equation}
\Psi\to\begin{cases}
e^{-i\left(\omega-\frac{qQ}{r_{+}}\right)r_{\ast}}\sim\left(r-r_{+}\right)^{-\frac{i}{2\kappa_{+}}\left(\omega-\frac{qQ}{r_{+}}\right)}, & r\to r_{+},\\
e^{i\left(\omega-\frac{qQ}{r_{c}}\right)r_{\ast}}\sim\left(r-r_{c}\right)^{-\frac{i}{2\kappa_{c}}\left(\omega-\frac{qQ}{r_{c}}\right)}, & r\to r_{c}.
\end{cases}\label{eq:boundary}
\end{equation}
Here $\kappa_{+}=\frac{1}{2f'(r_{+})}$ is the surface gravity on
the event horizon and $\kappa_{c}=-\frac{1}{2f'(r_{c})}$ the surface
gravity on the cosmological horizon. These asymptotic solution corresponds
to ingoing boundary condition near the event horizon and outgoing
boundary condition near the cosmological horizon. The system is dissipative
and the frequency of the perturbations will be the composition of
quasinormal modes (QNMs). With the specific boundary condition (\ref{eq:boundary}),
the radial equation (\ref{eq:radEq}) can be solved as an eigenvalue
problem. Only some discrete eigenfrequencies $\omega$ would satisfy
both the radial equation and boundary condition. One can write the
eigenfrequency as $\omega=\omega_{R}+i\omega_{I}$. When the imaginary
part $\omega_{I}>0$, the amplitude of perturbation will grow exponentially
and implies that the black hole is unstable at least in the linear
perturbation level.

\section{The instability of the 4D charged EGB-dS black hole }

The radial equation (\ref{eq:radEq}) is generally hard to solve analytically,
except in the regime where the frequency or the black hole size is
very small \cite{Brito2015,Li2019Hawking}. Many numerical method
for QNMs calculations were thus developed, such as WKB method, shooting
method, continued fraction method and Horowitz-Hubeny method \cite{Konoplya2011}.
Not all methods keeps high accuracy and efficiency in the charged
case \cite{Zhang2015}. In this work, we adopt the asymptotic iteration
method. Also, we testify our results from the asymptotic iteration methods with time evolution.

\subsection{The asymptotic iteration method (AIM)}

The asymptotic iteration method was used to solve the eigenvalues
of the homogeneous second order ordinary derivative functions \cite{Ciftci2005}.
Later it was used to look for the quasinormal modes of black holes
in asymptotic flat or (A)dS spacetime \cite{Cho2012}. Let us first
introduce an auxiliary variable
\begin{equation}
\xi=\frac{r-r_{+}}{r_{c}-r_{+}}.
\end{equation}
 It ranges from 0 to 1 as $r$ runs from the event horizon to the
cosmological horizon. The radial equation (\ref{eq:radEq}) then becomes
\begin{align}
0= & \frac{\partial^{2}\Psi}{\partial\xi^{2}}\left(\frac{f}{r_{c}-r_{+}}\right)^{2}+\frac{\partial\Psi}{\partial\xi}\frac{f\partial_{\xi}f}{\left(r_{c}-r_{+}\right)^{2}}\nonumber \\
 & +\left[\left(\omega-\frac{qQ}{(r_{c}-r_{+})\xi+r_{+}}\right)^{2}-f\left(\frac{l(l+1)+\left(\xi+\frac{r_{+}}{r_{c}-r_{+}}\right)\partial_{\xi}f}{\left[(r_{c}-r_{+})\xi+r_{+}\right]^{2}}\right)\right]\Psi.\label{eq:AIMeqxi}
\end{align}
The above equation is generally hard to solve analytically. We turn
to the numerical method. In terms of $\xi$, the asymptotic solution
near the horizons can be written as
\begin{align}
\Psi\to & \begin{cases}
\xi^{-\frac{i}{2\kappa_{+}}\left(\omega-\frac{qQ}{r_{+}}\right)}, & \xi\to0,\\
\left(\xi-1\right)^{-\frac{i}{2\kappa_{c}}\left(\omega-\frac{qQ}{r_{c}}\right)}, & \xi\to1.
\end{cases}\label{eq:AIMasymp}
\end{align}
 Then we can write the full solution of (\ref{eq:AIMeqxi}) satisfying
the asymptotic behavior (\ref{eq:AIMasymp}) in the following form
\begin{equation}
\Psi=  \xi^{-\frac{i}{2\kappa_{+}}\left(\omega-\frac{qQ}{r_{+}}\right)}\left(\xi-1\right)^{\frac{i}{2\kappa_{c}}\left(\omega-\frac{qQ}{r_{c}}\right)}\chi(\xi).
\end{equation}
Here $\chi(\xi)$ is a regular function of $\xi$ in range $(0,1)$
and obeys following homogeneous second order differential equation
\begin{equation}
\frac{\partial^{2}\chi}{\partial\xi^{2}}  =  \lambda_{0}(\xi)\frac{\partial\chi}{\partial\xi}+s_{0}(\xi)\chi,\label{eq:AIMeq}
\end{equation}
 in which the coefficients
 \begin{equation}
-\lambda_{0}(\xi)=  \frac{i\left(\omega-\frac{qQ}{r_{c}}\right)}{(\xi-1)\kappa_{c}}-\frac{i\left(\omega-\frac{qQ}{r_{+}}\right)}{\kappa_{+}\xi}+\frac{f'(\xi)}{f(\xi)},\\
\end{equation}
\begin{align}
-s_{0}(\xi)= & -\frac{\left(r_{c}-r_{+}\right)\left((\xi r_{c}-\xi r_{+}+r_{+})f'(\xi)+l(l+1)\left(r_{c}-r_{+}\right)\right)}{f(\xi)\left((\xi-1)r_{+}-\xi r_{c}\right){}^{2}}\\
 & -\frac{\left(\omega-\frac{qQ}{r_{c}}\right)\left(\frac{\omega r_{c}-qQ}{2\kappa_{c}r_{c}}+i\right)}{2(\xi-1)^{2}\kappa_{c}}+\frac{\left(\omega-\frac{qQ}{r_{+}}\right)\left(\frac{qQ-r_{+}\omega}{2\kappa_{+}r_{+}}+i\right)}{2\kappa_{+}\xi^{2}}+\frac{\left(\omega-\frac{qQ}{r_{+}}\right)\left(\omega-\frac{qQ}{r_{c}}\right)}{2\kappa_{+}(\xi-1)\xi\kappa_{c}}\nonumber \\
 & +\frac{if'(\xi)\left(\omega-\frac{qQ}{r_{c}}\right)}{2(\xi-1)\kappa_{c}f(\xi)}-\frac{if'(\xi)\left(\omega-\frac{qQ}{r_{+}}\right)}{2\kappa_{+}\xi f(\xi)}+\frac{\left(r_{c}-r_{+}\right){}^{2}\left(\omega-\frac{qQ}{\xi r_{c}-\xi r_{+}+r_{+}}\right){}^{2}}{f(\xi)^{2}}.\nonumber
\end{align}
The coefficients $\lambda_{0}(\xi)$ and $s_{0}(\xi)$ are regular
functions of $\xi$ in the interval $(0,1)$. Now using the same method
described in \cite{Zhang2020Super}, we can work out the quasinormal
modes. We vary the iteration times and the expansion point to ensure
the reliability of the results. The results are also checked by using
other auxiliary variables such as $\xi=1-\frac{r_{+}}{r}$ or $\xi=\left(1-\frac{r_{+}}{r}\right)/\left(1-\frac{r_{+}}{r_{c}}\right)$.
Except some extremal cases ($\Lambda\to0$ or the black hole becomes
extremal), they coincides well. Hereafter we set $q=1$ for convenience.

\subsection{The eigenfrequencies of the charged scalar perturbation }

\begin{figure}[htbp]
	\centering
	\includegraphics[scale=0.8]{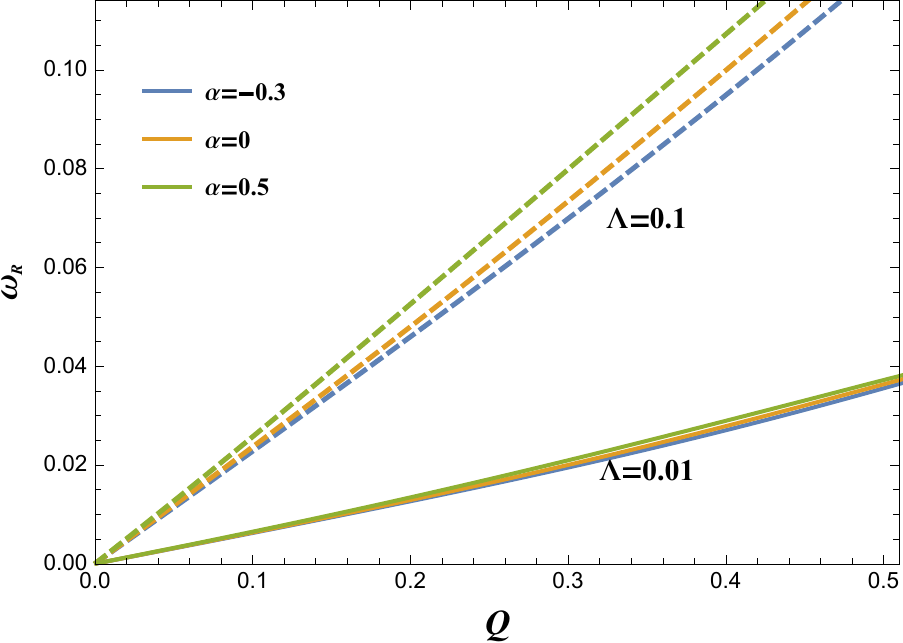}
	\includegraphics[scale=0.8]{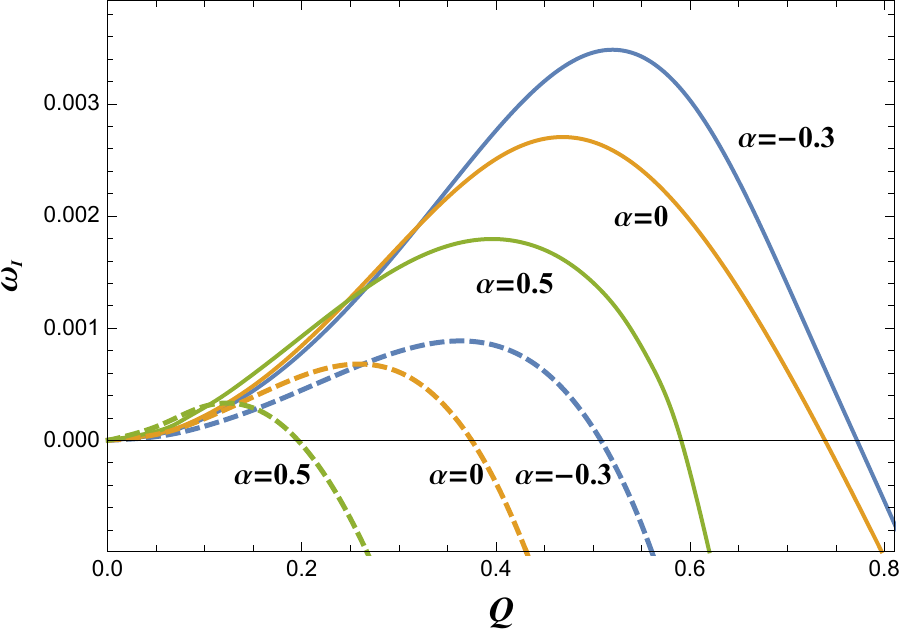}
	\caption{The real part (left) and imaginary part (right) of the fundamental modes of the QNMs when $l=0$. Solid lines for $\Lambda=0.01$, dashed lines for $\Lambda=0.1$}
	\label{fig:WQ}
\end{figure}

Let us first study the effects of the black hole charge $Q$ on the
fundamental modes of the QNMs. It is shown in Fig. \ref{fig:WQ}.
From the left panel, we see that $\omega_{R}$ increases with $Q$
almost linearly. The slope is larger for larger $\Lambda$ and $\alpha$.
In the right panel, $\omega_{I}$ increases with small $Q$ and then
decreases with larger $Q$. For large enough $Q$, the black hole
becomes stable. When $Q$ is small, positive $\alpha$ increases $\omega_{I}$
and negative $\alpha$ decreases $\omega_{I}$. This behavior is similar
to the case in asymptotic flat spacetime \cite{Zhang2020Super} .
However, when $Q$ is large, positive $\alpha$ decreases $\omega_{I}$
and negative $\alpha$ increases $\omega_{I}$. This is contrast with
the cases in asymptotic flat spacetime. It implies that the positive
GB coupling constant can make the black hole more unstable when the
black hole charge $Q$ is small and more stable when $Q$ is large.
Note further that no matter what values $\Lambda,\alpha$ take, when
$Q\to0$, both $\omega_{R}$ and $\omega_{I}$ tend to 0 from above.
This implies the weakly charged black hole in dS spacetime is always
unstable. The existence of $\alpha$ does not change this phenomenon
qualitatively.
\begin{figure}[h]
	\centering
	\includegraphics[scale=0.56]{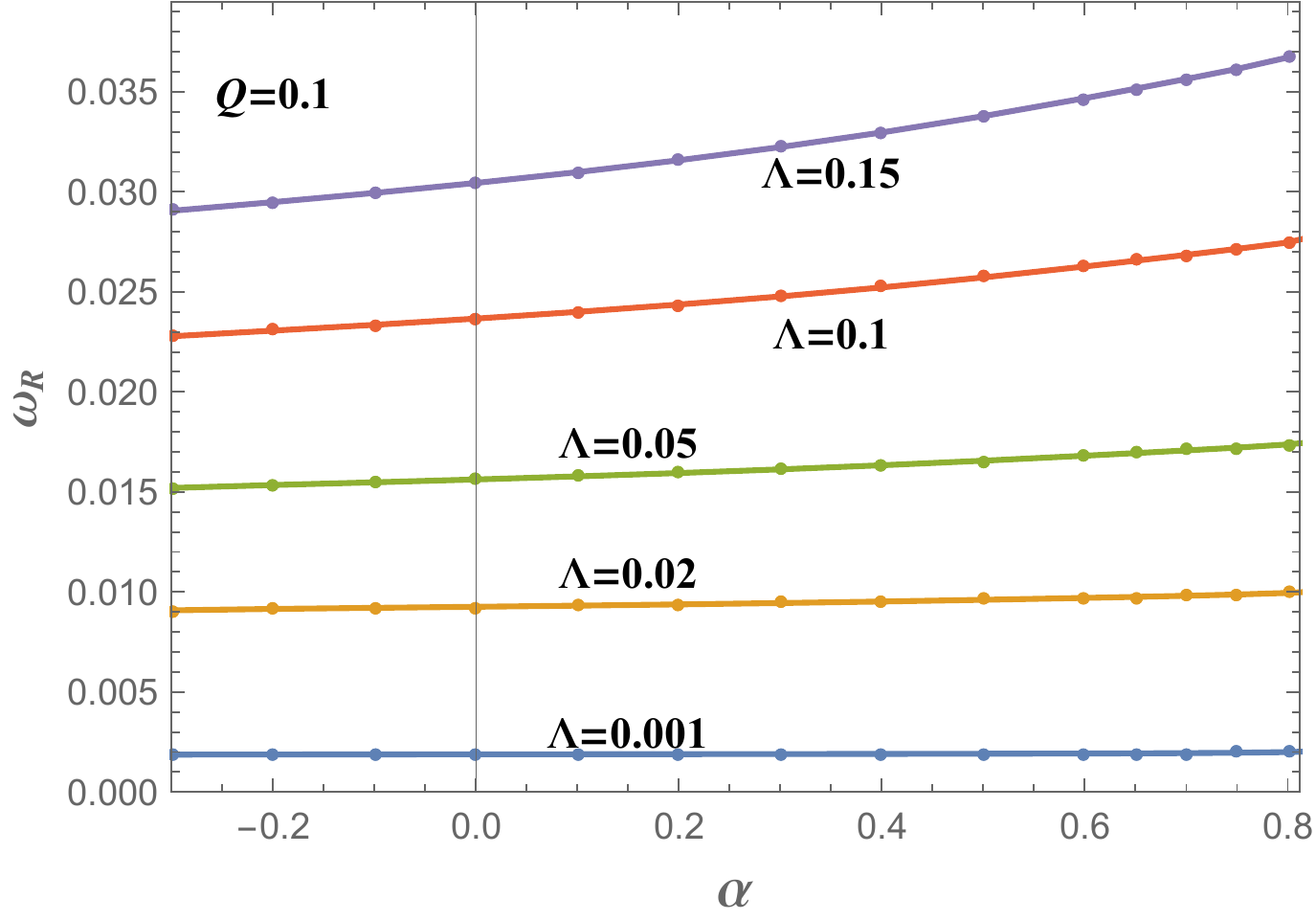}
	\includegraphics[scale=0.56]{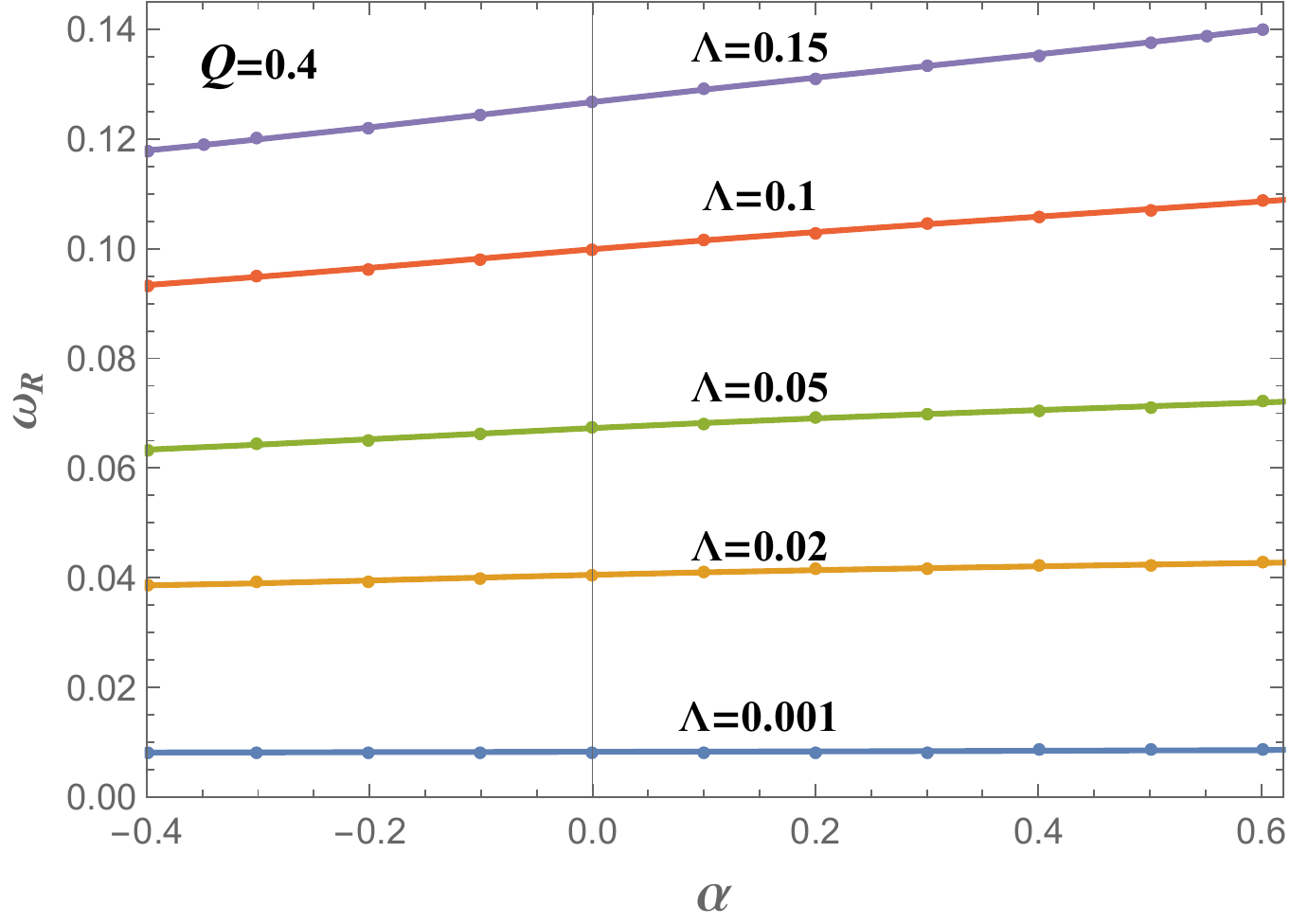}
	\caption{The real part of the fundamental modes of the QNMs when $l=0$. Left panel for $Q=0.1$, right panel for $Q=0.4$.}
	\label{fig:QAlphaWR}
\end{figure}

Now we study the effects of $\alpha$ on the fundamental modes in
detail. Fig. \ref{fig:QAlphaWR} shows the real part of the fundamental
modes $\omega_{R}$ when $Q=0.1$ and $Q=0.4$. For fixed $Q$, the
real part $\omega_{R}$ increases almost linearly with $\alpha$ and
$\Lambda$. Combining the results from Fig. \ref{fig:WQ}, we conclude
that $\omega_{R}\propto\alpha\Lambda Q$.

\begin{figure}[h]
	\centering
	\includegraphics[scale=0.56]{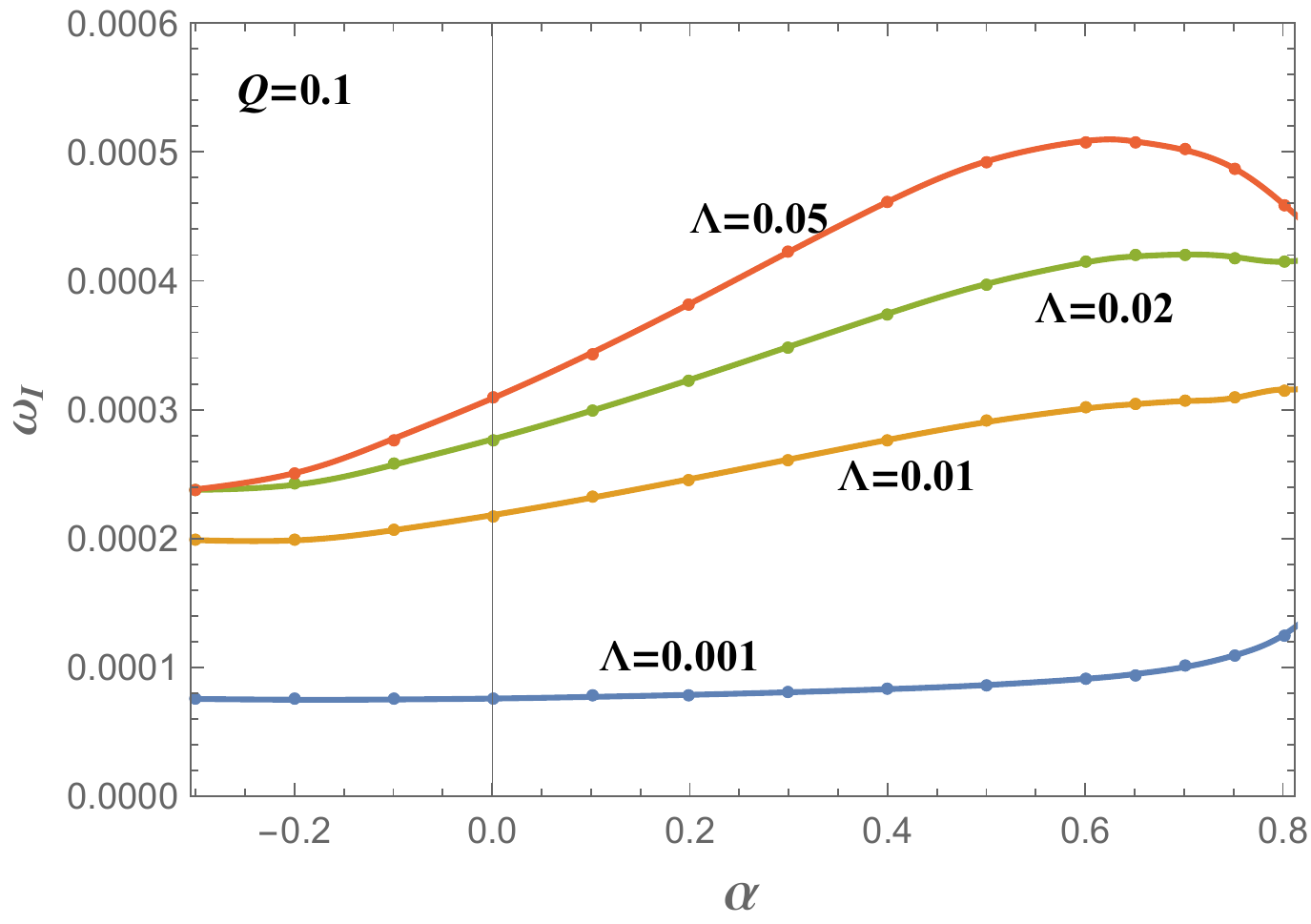}  \includegraphics[scale=0.56]{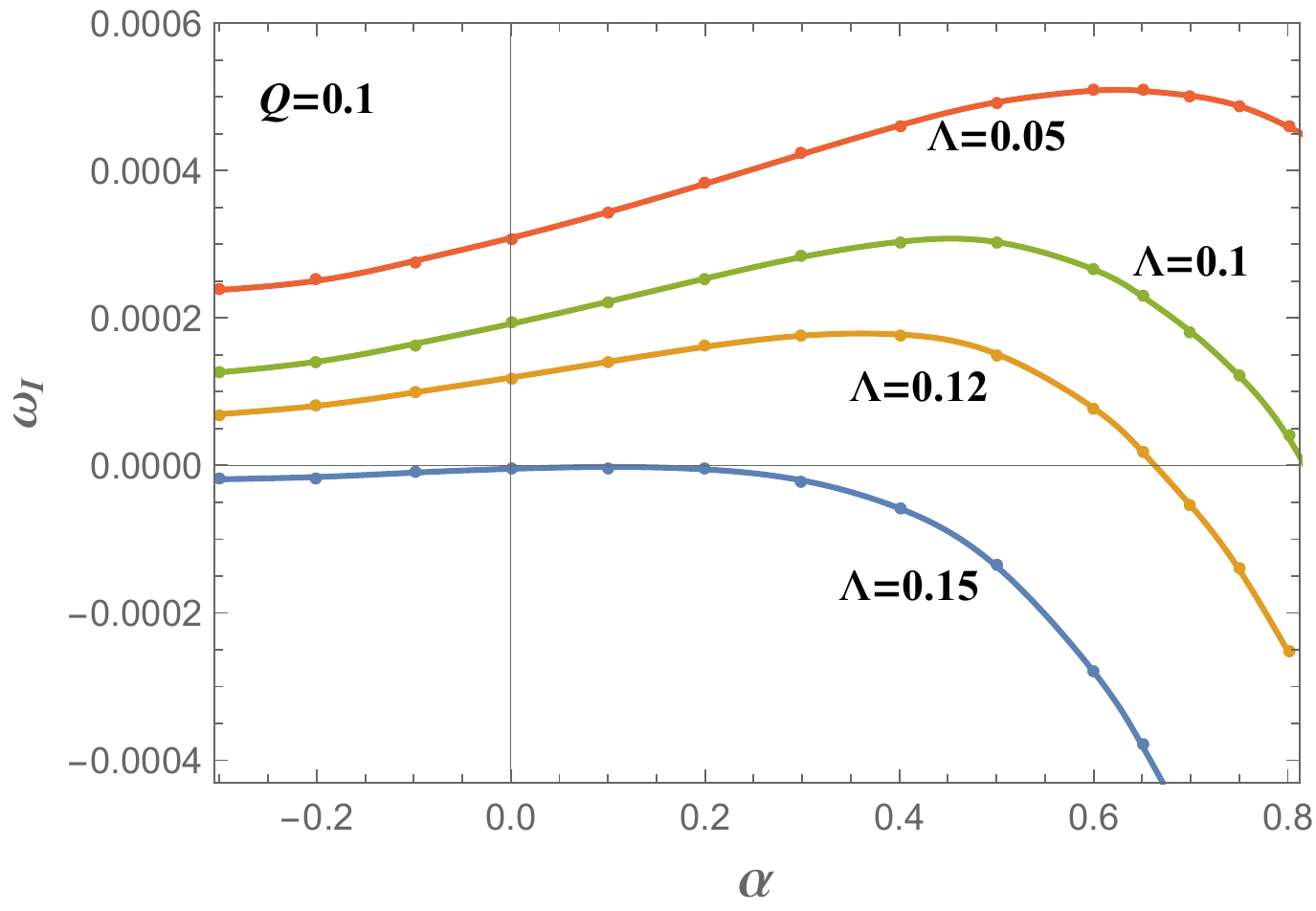}
	\includegraphics[scale=0.56]{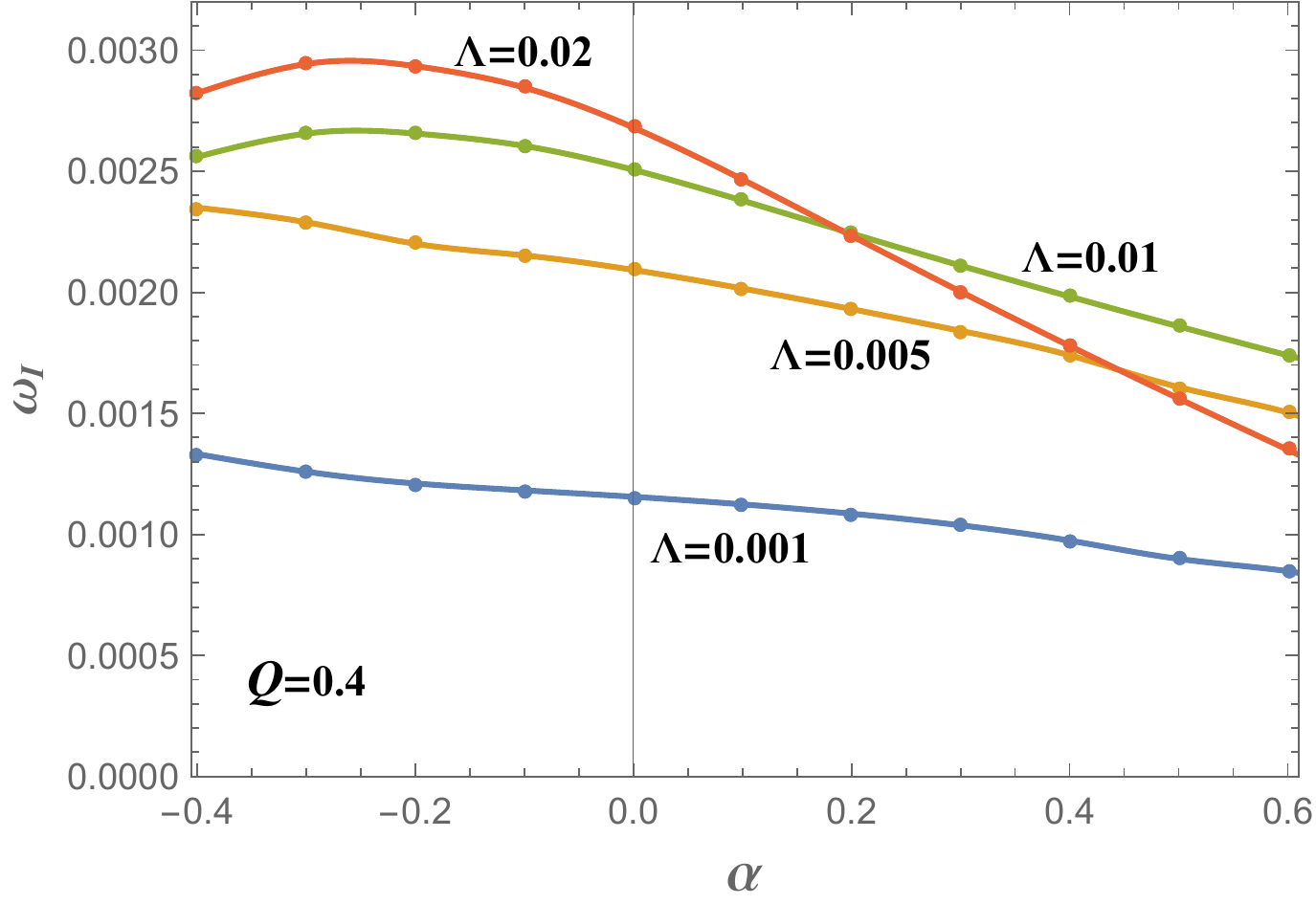}  \includegraphics[scale=0.56]{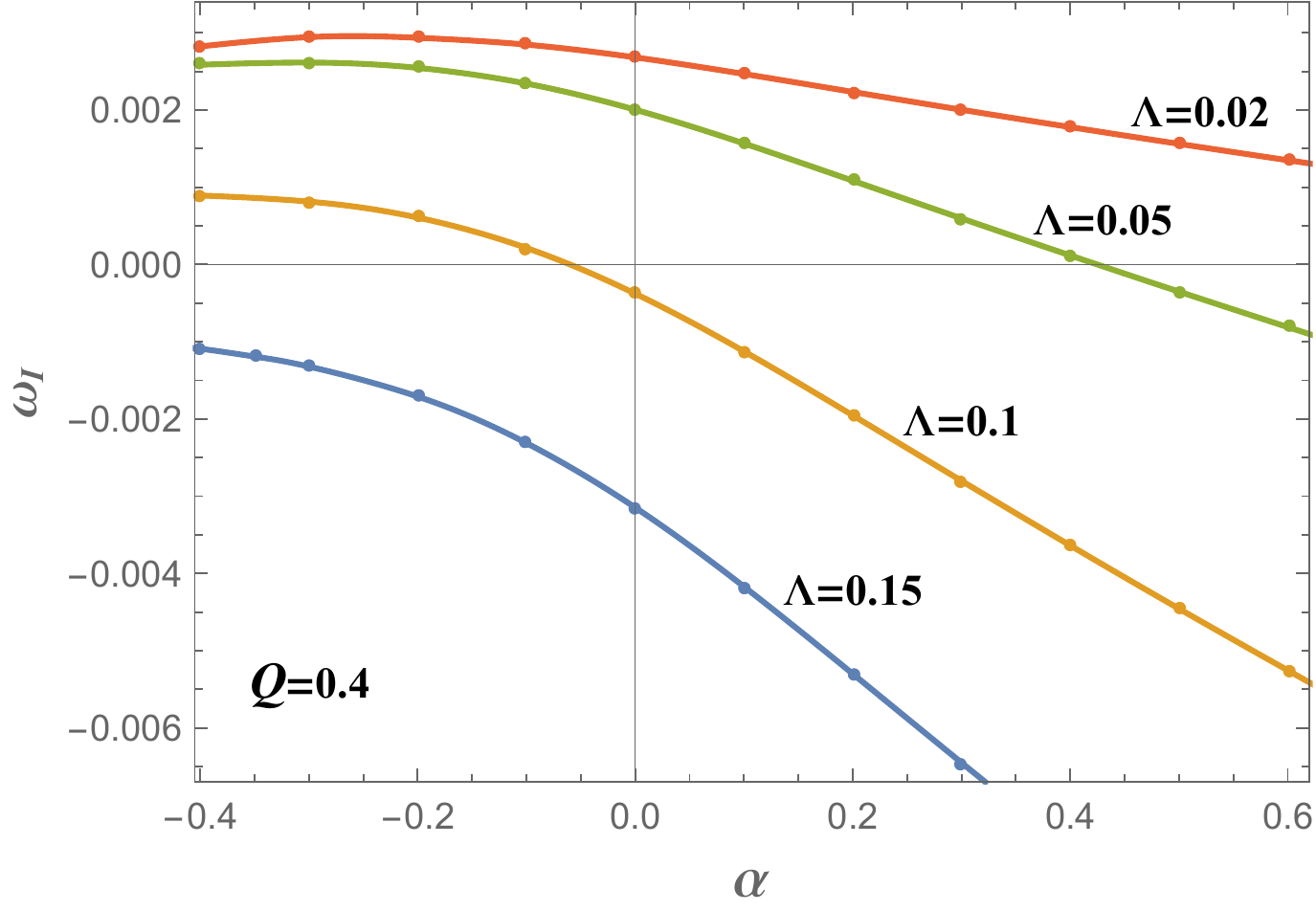}
	\caption{The imaginary part of the fundamental modes of the QNMs when $l=0$. The upper panel for $Q=0.1$, lower panel for $Q=0.4$. Left panel shows the cases for small $\Lambda$, the right panel for larger $\Lambda$.}
	\label{fig:QAlphaWI}
\end{figure}

The behavior of the imaginary part of the fundamental modes $\omega_{I}$
is more interesting. In Fig. \ref{fig:QAlphaWI} we show the cases
when $Q=0.1$ and $Q=0.4$. From the left and right panels, we see
that $\omega_{I}$ increases with small $\Lambda$ and decreases with
larger $\Lambda$. The effects of $\alpha$ on $\omega_{I}$ is subtle
and relevant to the $\Lambda$ and $Q$. When both $Q$ and $\Lambda$
are small (left upper panel), $\omega_{I}$ increases with $\alpha$.
For small $Q$ and larger $\Lambda$ (right upper panel), $\omega_{I}$
increases with $\alpha$ first and then decreases with $\alpha$.
For larger $Q$ (lower panels), $\omega_{I}$ decreases with $\alpha$.
This phenomenon is very different with the case in asymptotic flat
spacetime \cite{Zhang2020Super}, where $\alpha$ roughly increases
$\omega_{I}$. This implies that the positive GB coupling constant
$\alpha$ can suppress the instability of black hole. As $\alpha$
increase, it can even change the qualitative behavior of black hole
under perturbations and render the unstable black hole stable.

The instability we found here is very reminiscent of superradiance.
However, the case is subtle here. With the similar method used in
\cite{Zhang2020Super,Konoplya2014b}, one can show that superradiance
occurs only when
\begin{equation}
\frac{qQ}{r_{+}}>\omega>\frac{qQ}{r_{c}}.\label{eq:SuperCond}
\end{equation}
We see that some modes satisfying (\ref{eq:SuperCond}) are unstable,
while there are indeed some modes satisfying (\ref{eq:SuperCond})
are stable. See Table \ref{tab:The-fundamental-modes} for evidence.
In fact, as shown in \cite{Zhang2020Super,Konoplya2014b}, the superradiant
condition is the necessary but not sufficient condition for instability.
The precise mechanism of the instability found here need more studies and will be addressed in section 5.
\begin{center}
\begin{table}[h]
\begin{centering}
\begin{tabular}{|c|c|c|c|c|}
\hline
$\alpha$ & $\frac{qQ}{r_{+}}$ & $\frac{qQ}{r_{c}}$ & $\omega$ (AIM) & $\omega_I$ (time domain)\tabularnewline
\hline
{\small{}0.5} & {\small{}0.1} & {\small{}0.0244042} & {\small{}0.0290311+0.0001514$i$} & {\small{0.0001515}}\tabularnewline
{\small{}0.6} & {\small{}0.1} & {\small{}0.0248557} & {\small{}0.0297114+0.0000779$i$} & {\small{0.0000780}}\tabularnewline
{\small{}0.65} & {\small{}0.1} & {\small{}0.0250987} & {\small{}0.0300692+0.0000191$i$} & {\small{}0.0000192}\tabularnewline
{\small{}0.7} & {\small{}0.1} & {\small{}0.0253547} & {\small{}0.0304355--0.0000541$i$} & {\small{}-0.0000547}\tabularnewline
{\small{}0.75} & {\small{}0.1} & {\small{}0.0256252} & {\small{}0.0308094--0.0001400$i$} & {\small{}-0.0001390}\tabularnewline
\hline
\end{tabular}
\par\end{centering}
\caption{\label{tab:The-fundamental-modes}
{\small{}The fundamental modes when
$Q=0.1,\Lambda=0.12$ and $l=0$, which corresponds to the orange line in the upper right panel of Fig. \ref{fig:QAlphaWI}. The last column is the imaginary part of the frequency extracted from the time evolution.}}
\end{table}
\par\end{center}

We also directly compute the time-evolution of the perturbation field $ \psi $ to further reveal the instability of the 4D charged EGB-dS black hole. For time evolution, the Schr\"odinger-like equation becomes,
\begin{equation}\label{eq:eomvsl}
  -\frac{\partial^{2} \Psi}{\partial t^{2}} - \frac{2 i q Q}{r} \frac{\partial \Psi}{\partial t}+\frac{\partial^{2} \Psi}{\partial r_{*}^{2}}-V(r) \Psi=0,
\end{equation}
In order to compute the time evolution of $\psi$ we adopt the discretization introduced in \cite{Zhu2014}. The reliability of this numerical method can be verified by the convergence of computations when increasing the sampling density. We impose the following initial profile,
\begin{equation}\label{eq:initial}
  \left\{
  \begin{aligned}
    \Psi(r_*,t) & =0,                                          & t<0, \\
    \Psi(r_*,t) & =\exp\left[{-\frac{(r_*-a)^2}{2b^2}}\right], & t=0.
  \end{aligned}
  \right.
\end{equation}
The discretization of equation \eqref{eq:eomvsl} is implemented in $(r_*,t)$ plane. Also, we fix $\Delta t/\Delta r_* = 0.5$ in order to satisfy the von Newmann stability conditions. Unlike some other analytical models where the $r_*$ can be solved analytically, here the $r_*(r)$ can only be obtained numerically. The $ r_*(r) $ diverges as $ \lim_{r\to r_+} \to -\infty $ and $ \lim_{r\to r_c} \to \infty $, hence the error of the numerical $ r_*(r) $ can be very large at the near horizon region. Therefore, we introduce a cutoff $ \epsilon $ solve the $r_*(r)$ relation by solving the below system,
\begin{equation}\label{eq:rstartvsr}
  r_*'(r) = 1/f(r),\,r_*(r_h + \epsilon) = 0, \, \text{with}\, r\in [r_h + \epsilon,r_c - \epsilon],
\end{equation}
Usually the cutoff $\epsilon$ should not be too small, otherwise it leads to significant error of the resultant $ r_*(r) $.
\begin{figure}[h]
	\centering
	\includegraphics[width=0.45\textwidth]{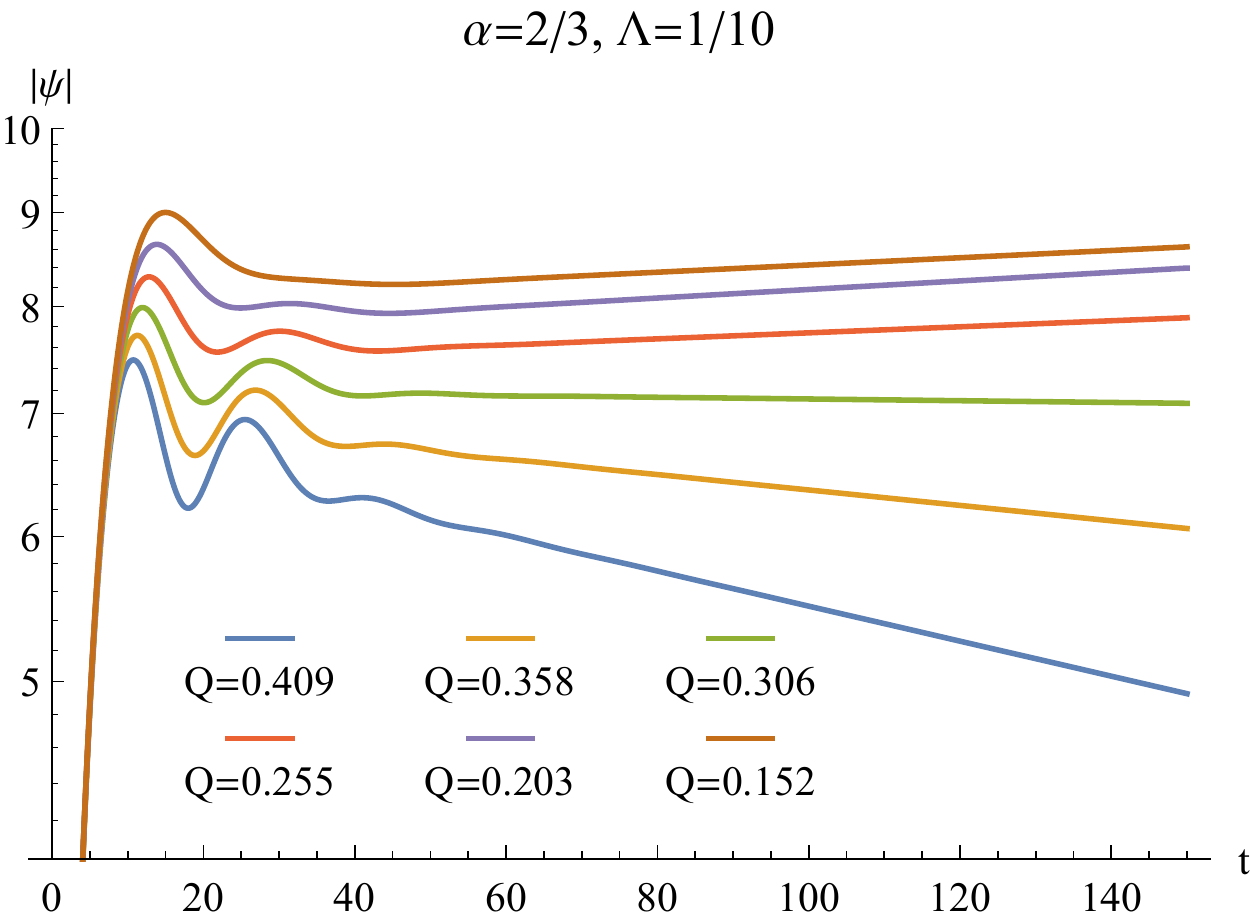}
	\includegraphics[width=0.45\textwidth]{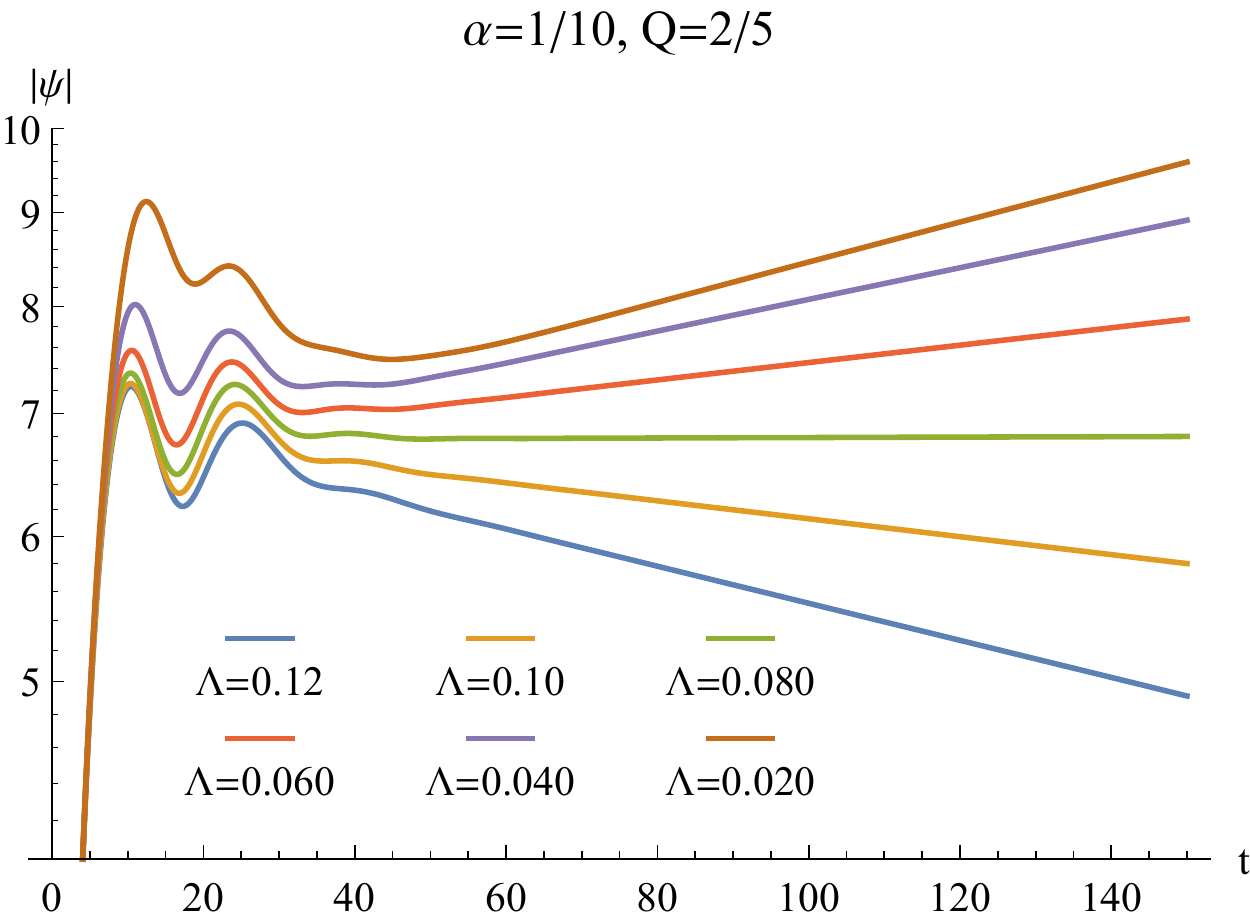}
	\caption{Left panel: the time evolution of the $ |\psi(r_* = 88.4216,t)| $ at $ \alpha = 2/3,\,\Lambda = 1/10$, where each curve corresponds to different values of $ Q $. Right panel: the time evolution of the $ |\psi(r_* = 88.4216,t)| $ at $ \alpha = 1/10,\,Q = 2/5 $, where each curve corresponds to different values of $ \Lambda $. For both plots we fixed $ a = 88.4216,\,b = 1/10 $.}
	\label{fig:plotshow1}
\end{figure}

In order to obtain the late time evolution of the perturbation, we need to solve a large range or $r_*$. Since $1/f$ tends to diverge when $r\to r_h$ and $r\to r_c$, in near horizon region the $r_*(r)$ can be extracted analytically. A direct solution is to expand $1/f(r)$ with respect to $r_h$, because the horizon requires that $1/f(r) \sim 1/(r-r_h) $. However, this direct expansion can lead to very large numerical error. A better solution is to expand $f(r)$ with respect to $r_h$, and then obtain the expansion of $1/f$ in terms of the expansion coefficients of the $f(r)$. After solving the analytical expansion coefficients, one may glue the analytical expansion in the near horizon region and the numerical solution of the $r_*(r)$ in $ r\in [r_h + \epsilon,r_c - \epsilon] $. In this way, one may obtain a very large range of $r_*$ and a long-time evolution can be realized.

We show two examples of $ |\psi(t)| $ in log plot in Fig. \ref{fig:plotshow1}, from which we can see that $ \ln |\psi| $ has linear dependence on $ t $ for late time evolution. From the left panel of Fig. \ref{fig:plotshow1}, when $ Q $ is small the system is unstable ($ \ln |\psi| $ linearly decreases with $ t $), while for large values of $ Q $ the system becomes stable ($ \ln |\psi| $ linearly grows with $ t $). This is in accordance with previous results of the frequency analysis (see the right panel of Fig. \ref{fig:WQ}). From the right panel of Fig. \ref{fig:plotshow1} we see that when $ \Lambda $ is small the system is unstable ($ \ln |\psi| $ linearly increases with $ t $), while for large values of $ \Lambda $ the system becomes stable ($ \ln |\psi| $ linearly decreases with $ t $). This is in accordance with previous results of the frequency analysis (see the bottom right panel of Fig. \ref{fig:QAlphaWI}).

It is also important to verify the validity of the asymptotic iteration method with the time evolution. Specifically, in the time evolution computation, we can extract the $ \omega_R $ by computing the $ \partial_t\ln(|\psi|) $ at late time and compare that with those from the frequency analysis. We provide the comparison in Table \ref{tab:The-fundamental-modes} (see the last two columns), from which we can see that all the time evolution results perfectly matches with those of the asymptotic iteration method.

\begin{figure}[h]
\centering
\includegraphics[scale=1]{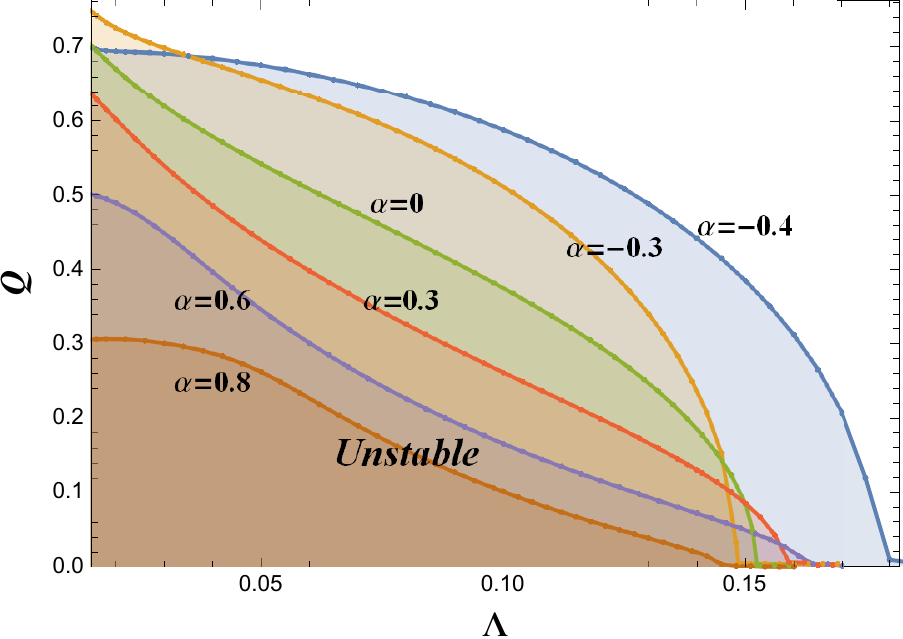}
\caption{\label{fig:UnstableRegion}
The unstable region of the charged
EGB-dS black hole under charged scalar perturbation. The shadows under
constant $\alpha$ lines are corresponding unstable regions. We fix
$q=1$ here.}
\end{figure}

Finally, we show the unstable region of the charged EGB-dS black hole
under charged scalar perturbation in Fig. \ref{fig:UnstableRegion}.
The black hole is unstable only when $\Lambda$ and $Q$ are not too
large. As $\Lambda$ or $Q$ increase, the black hole becomes less
unstable. For positive $\alpha$, the unstable region shrinks. For
negative $\alpha$, the unstable region enlarges. Although we do not
show the results when $\Lambda\to0$ due to the limitation of our
numerical method, we can expect that there should be sudden drops
since we have found that there is no instability for asymptotic flat
charged EGB black hole under charged scalar perturbation \cite{Zhang2020Super}.
This phenomenon was also disclosed for the RN-dS black hole in \cite{Konoplya2014b}.
\begin{figure}[h]
	\centering
	\includegraphics[scale=0.56]{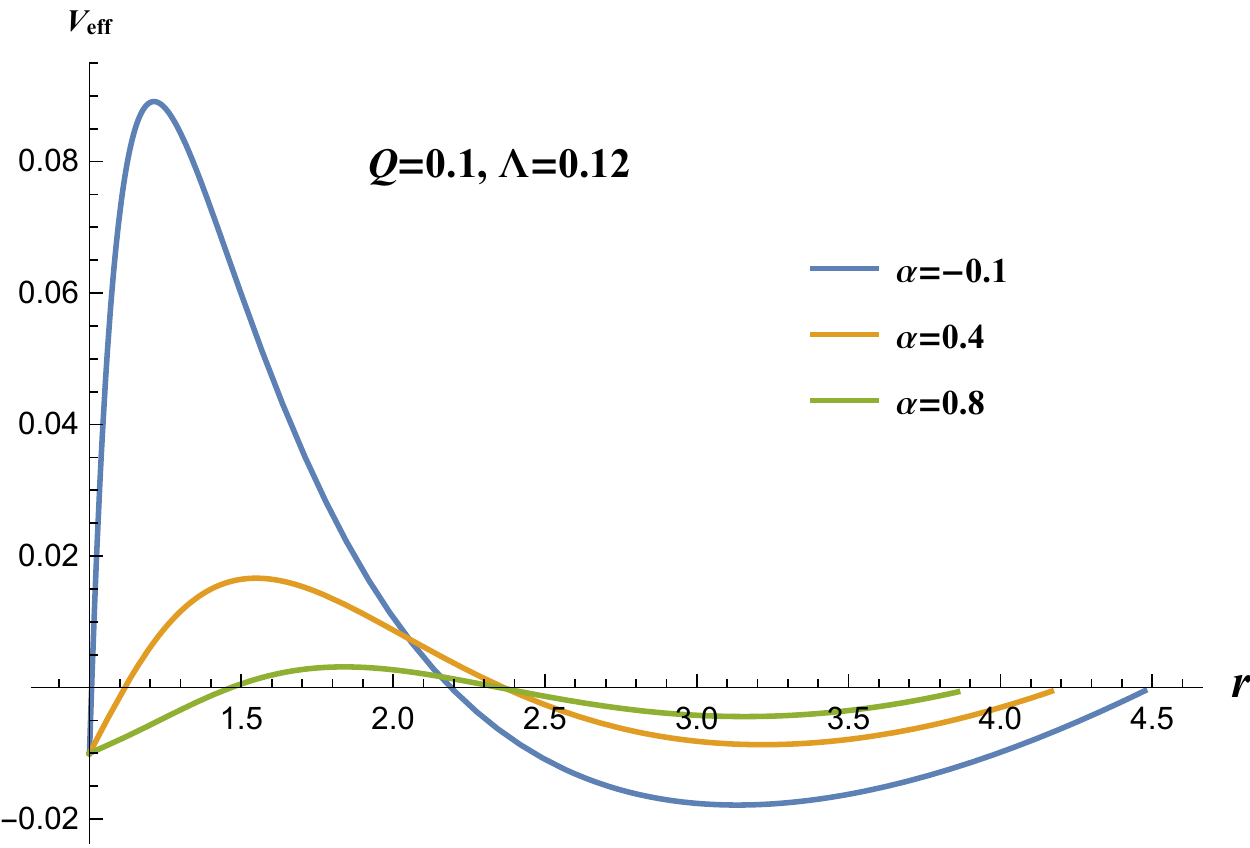}
	\includegraphics[scale=0.56]{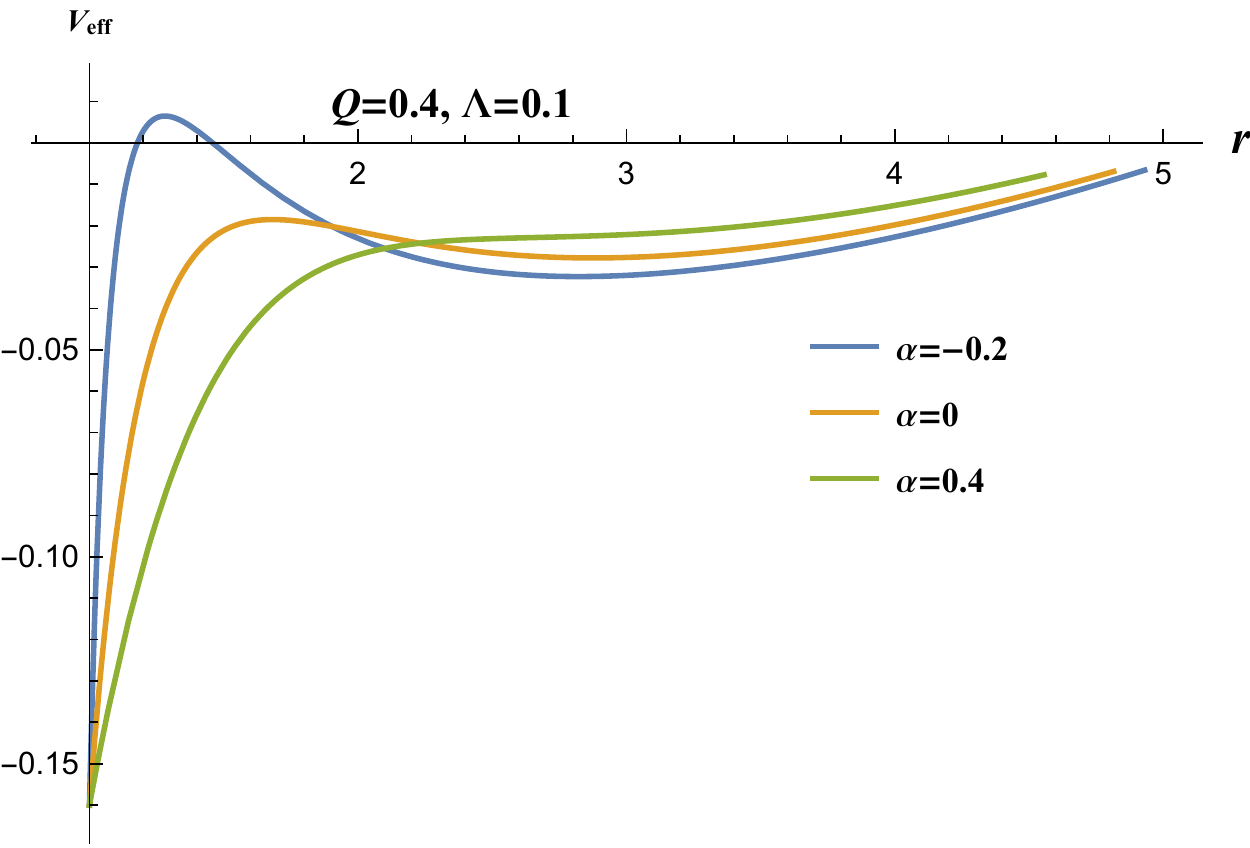}
	\caption{The effective potential when $l=0$. They correspond to the orange line in the upper right panel and orange line in the lower right panel of Fig \ref{fig:QAlphaWI}, respectively.}
	\label{fig:VeffL0}
\end{figure}

Now let us take a look at the effective potential when $l=0$. In
the left panel of Fig. \ref{fig:VeffL0}, we see that there is a negative
potential well between $r_{+}$ and $r_{c}$. This potential well
is the key point for the occurrence of instability. However, the negative
effective potential does not guarantee $\omega_{I}>0$. For example,
the case with $\alpha=0.8$ has a negative potential well, but the
corresponding $\omega_{I}<0$. Thus the existence of a negative potential
well can be view as the necessary but not sufficient condition for
the instability \cite{Konoplya2012}. In the right panel of Fig. \ref{fig:VeffL0},
the potential well disappears for larger $\alpha$. The perturbation
wave can be easily absorbed by the black hole and the corresponding
background becomes stable under charged scalar perturbations. Note
that the positive cosmological constant is crucial here to creating
the necessary potential well.

\begin{figure}[h]
	\centering
	\includegraphics[width=0.45 \textwidth]{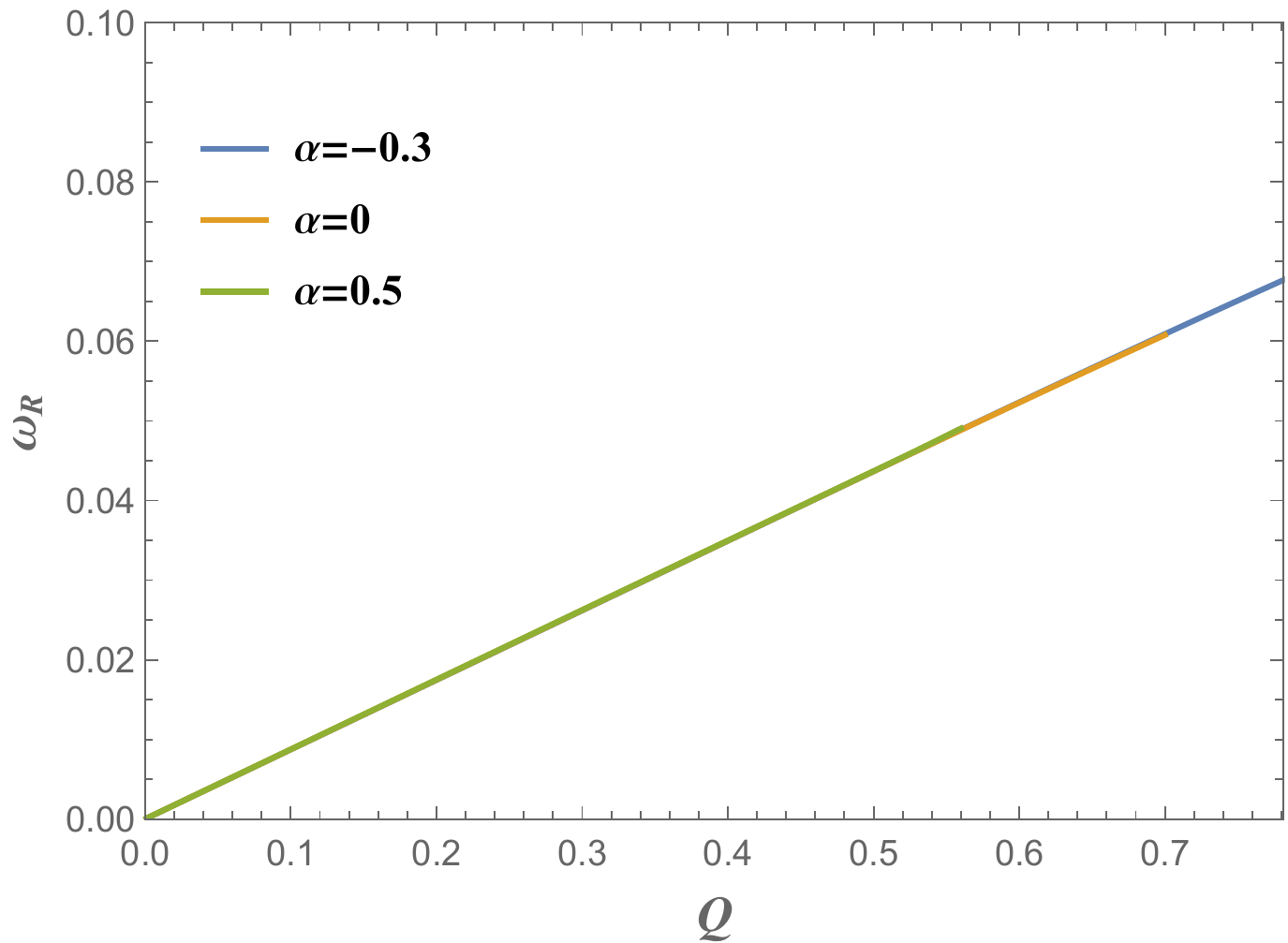}
	\includegraphics[width=0.45 \textwidth]{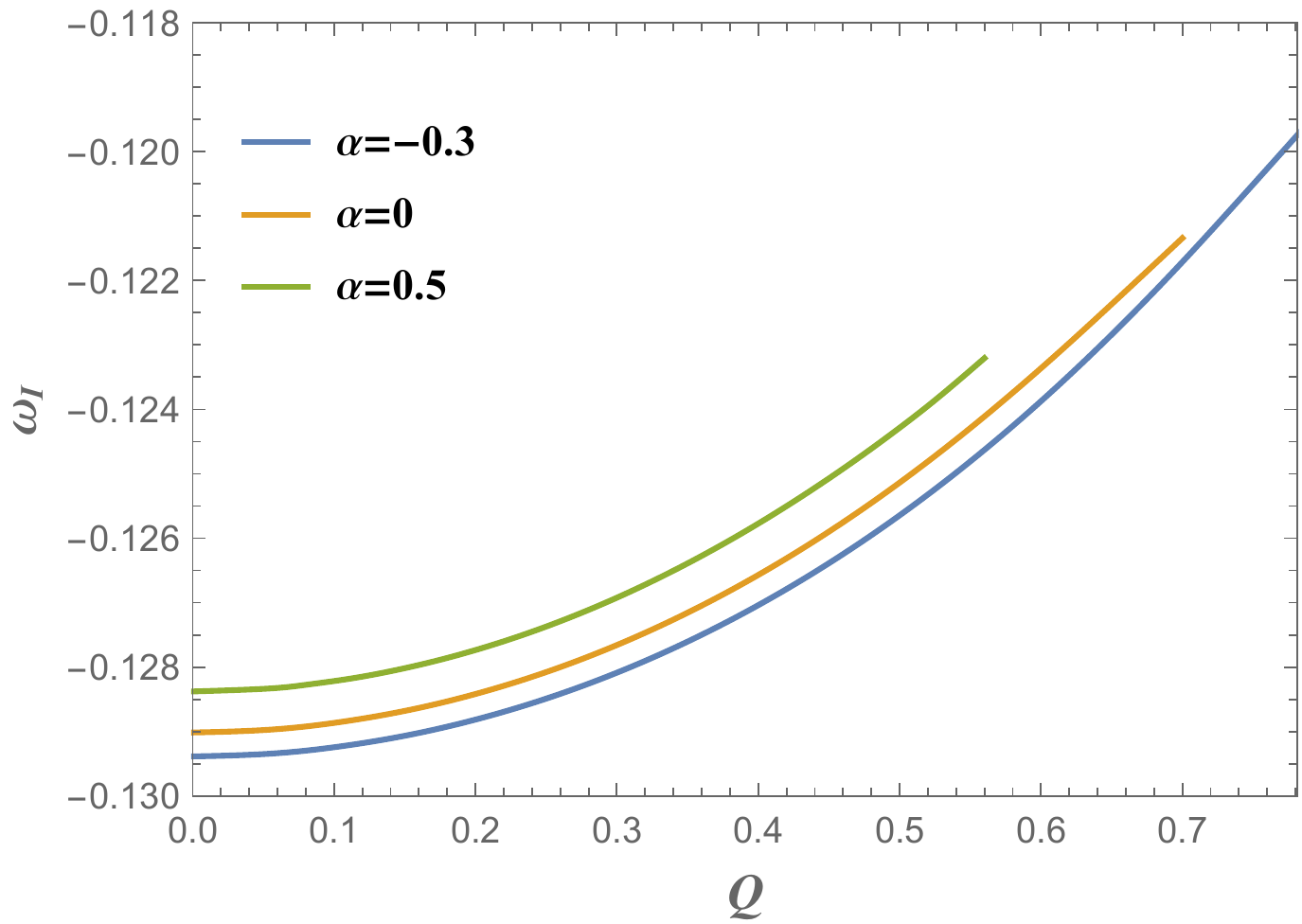}
	\caption{The real part (left) and imaginary part (right) of the fundamental modes when $l=1$. We fix $\Lambda=0.05$ here. The case with other $\Lambda$ are similar.}
	\label{fig:WQL1}
\end{figure}

Now we consider the eigenfrequencies of the charged scalar perturbation
when $l=1$. We show the fundamental modes in Fig. \ref{fig:WQL1}.
The left panel shows that there is still $\omega_{R}\propto Q$ for
higher $l$. But $\alpha$ changes $\omega_{R}$ little. All the fundamental
modes has $\omega_{R}<\frac{qQ}{r_{c}}$ that live beyond the superradiant
condition. The left panel is different from that in Fig. \ref{fig:WQ}.
Here $\omega_{I}$ increases with $Q$ and $\alpha$ monotonically.
Note that all the modes are stable now.

\begin{figure}[h]
	\centering
	\includegraphics[scale=0.6]{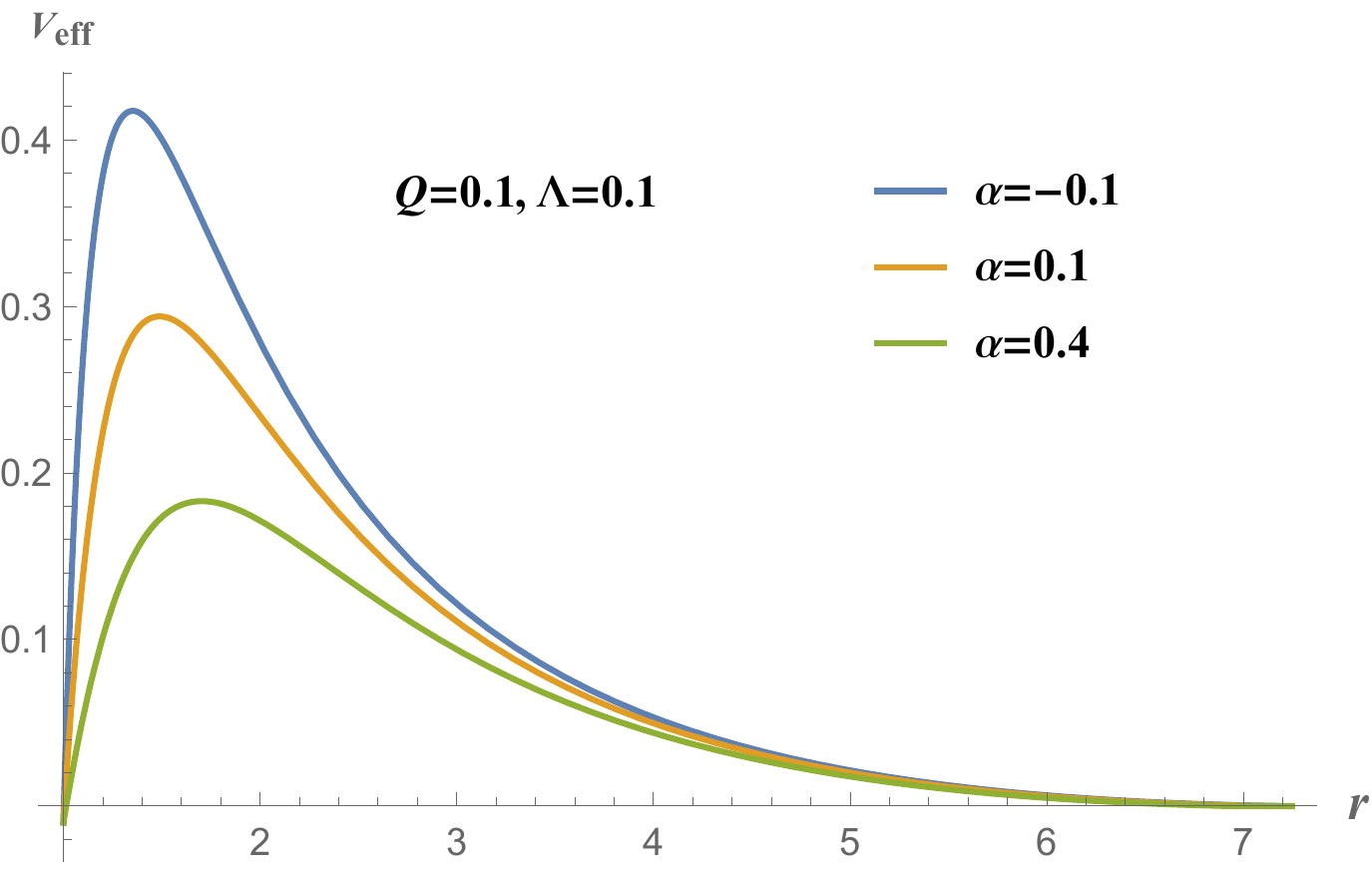}
	\caption{The effective potential when $l=1$. The cases with other $Q,\Lambda$ are similar.}
	\label{fig:VeffL1}
\end{figure}

The stability of the higher $l$ can be explained from the effective
potential, as shown in Fig. \ref{fig:VeffL1}. There is only one potential
barrier and not potential well to accumulate the energy to trigger
the instability.

\begin{figure}[htbp]
	\centering
	\includegraphics[width=0.45\textwidth]{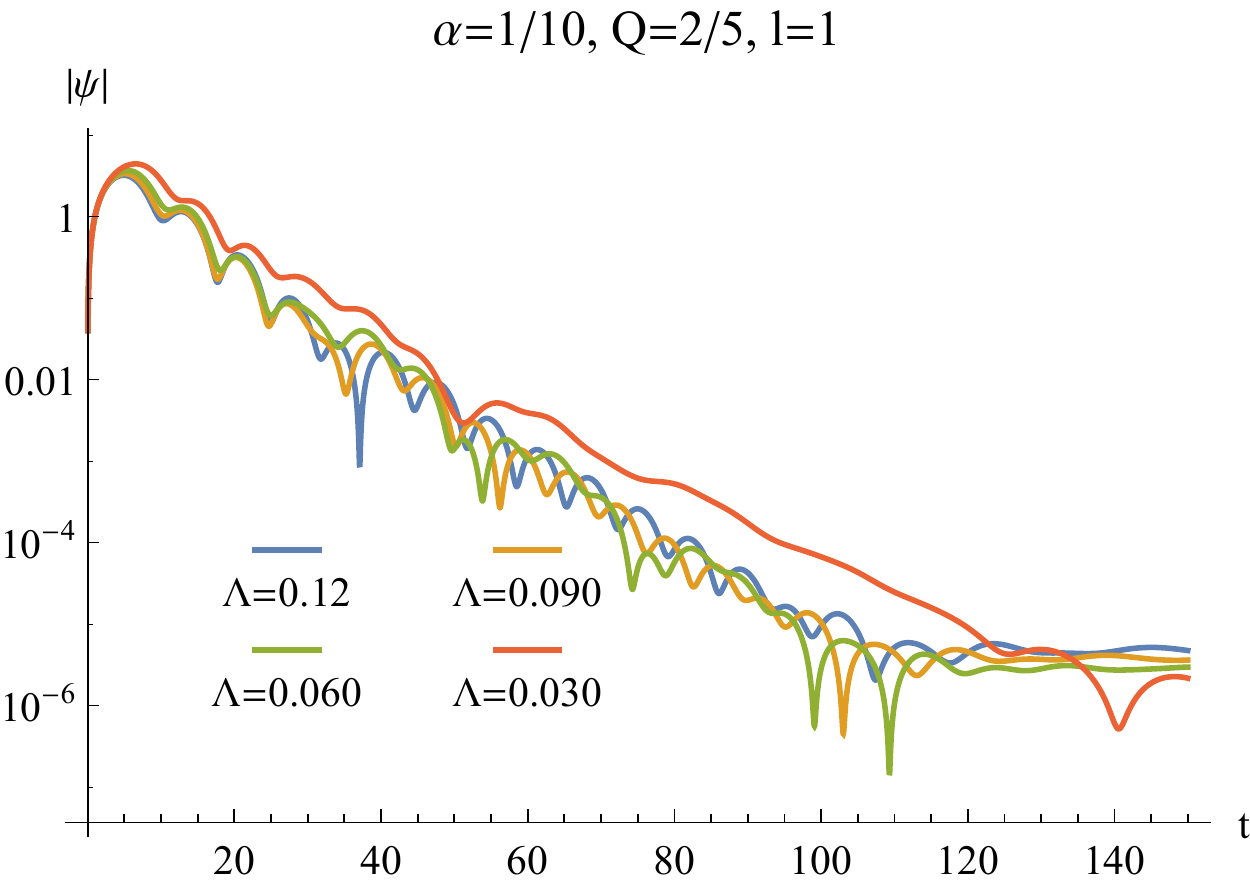}
	\includegraphics[width=0.45\textwidth]{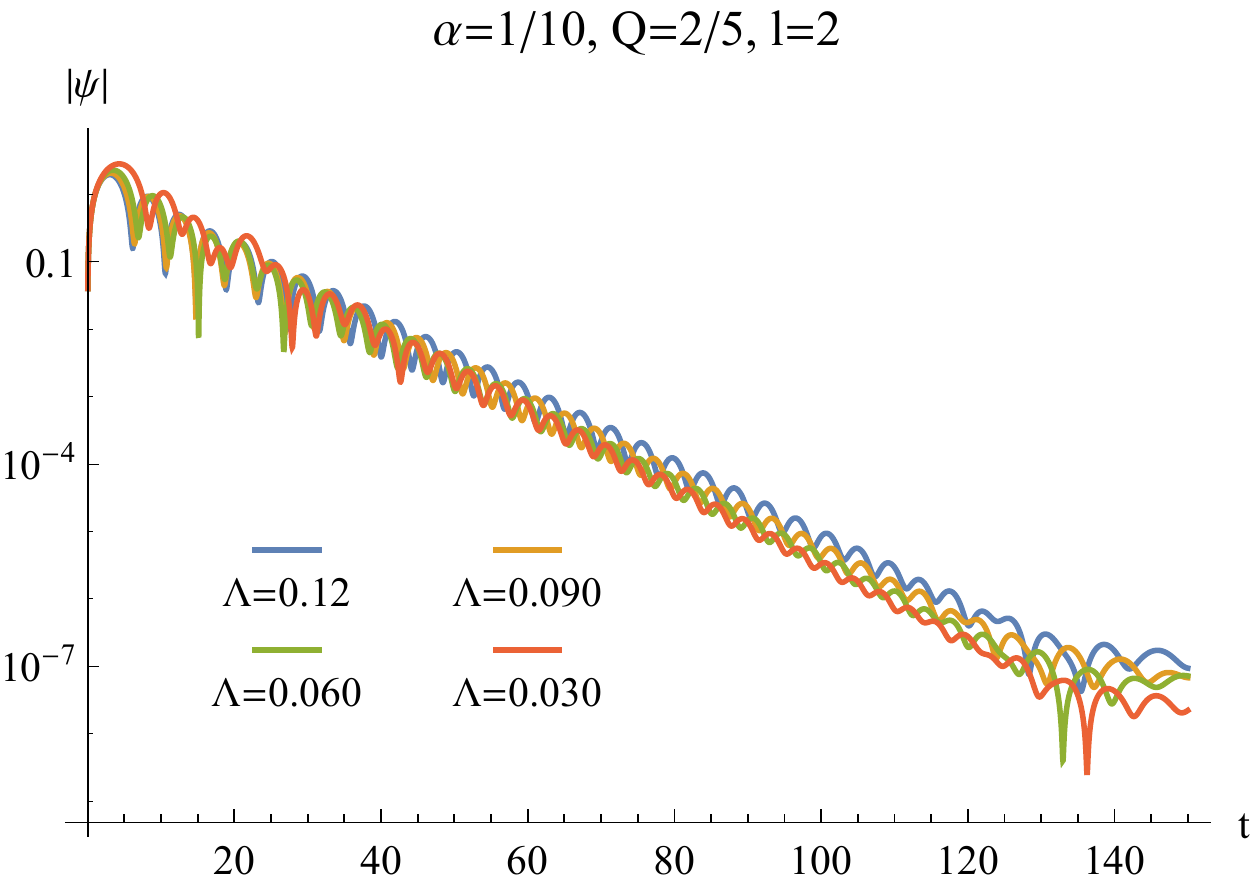}
	\caption{The time evolution of the $ |\psi(r_* = 88.4216,t)| $ in log plot at $ \alpha = 1/10, \, Q = 2/5 ,\, a = 88.4216, \, b = 1/10$ and $ q=1 $, where each curve in both panels corresponds to different values of $ \Lambda $. The left and the right panel corresponds to $ l=1 $ and $ l=2 $, respectively.}
	\label{fig:lneq0}
\end{figure}

We also provide the detailed time evolution for $ l \neq 0 $ in Fig. \ref{fig:lneq0}. From these two panels we see that the perturbations indeed decays in long time evolution and larger $ l $ leads to more significant decay of $ \psi $, and hence no instability occurs for higher values of $ l $.

\section{Summary and discussion}

We studied the instability of charged 4D EGB-dS black hole under the
charged massless scalar perturbation. This instability satisfies the
superradiant condition. However, not all the modes satisfying the
superradiant condition are unstable. The precise mechanism for the
this instability is not well understood. But the positive cosmological
constant $\Lambda$ should play a crucial role. The instability occurs
when the cosmological constant is small. This is reminiscent of the
Gregory-Laflamme instability \cite{Gregory1993} since here exists
hierarchy between the black hole event horizon and cosmological horizon.
The instability here is different with the ``$ \Lambda $ instability'' found in
\cite{Cuyubamba2020,Konoplya2008,Li2019RNdS} which occurs when the
black hole charge and the cosmological constant are large.

We analyzed this instability from the viewpoint of the effective potential.
Higher $l$ has only one potential barrier beyond the event horizon.
The perturbation dissipates and leads no instability. The monopole
$l=0$ has a negative potential well between the event horizon and
cosmological horizon, which can accumulate the energy to trigger the
instability. But the negative potential well is just the necessary
but not sufficient condition for the instability.

Unlike the cases in asymptotic flat spacetime, the effects of the
GB coupling constant $\alpha$ on the perturbation is relevant to
the black hole charge and cosmological constant. It makes the unstable
black hole more unstable when both the black hole charge and cosmological
constant are small, and makes the stable black hole more stable when
the black hole charge is large. The weakly charged black hole in dS
spacetime is always unstable. The existence of $\alpha$ does not
change this phenomenon qualitatively. However, $\alpha$ can change
the qualitative behavior when the black hole charge is large, and
make the unstable black hole stable. We show that the unstable region
of ($\Lambda,Q$) shrinks with positive $\alpha$ can enlarges with
negative $\alpha$. Unfortunately, we do not get the  accurate enough results
when $\Lambda \to 0$  due to the limitation of our numerical method. The case when black hole becomes extremal is also beyond the effectiveness of this method.
The stability of extremal black hole may be very different with   that of non-extremal black hole. There is ``horizon instability'' universally \cite{Aretakis2011}.  We leave them for further study.

We point out several topics worthy of further investigations. The stability of massive perturbation should be explored in detail to reveal how mass term affects the stability of the charged perturbation. The stability of the 4D charged Einstein-Gauss-Bonnet anti de-Sitter are also definitely interesting to find out. We plan to explore these directions in near future.

\section{Acknowledgments}

We thanks Peng-Cheng Li, Minyong Guo and Shao-Jun Zhang for helpful discussions. Peng Liu would like to thank Yun-Ha Zha for her kind encouragement during this work.
 Peng Liu is supported by the Natural Science Foundation of China under Grant No. 11847055, 11905083.
Chao Niu is supported by the Natural Science Foundation of China under Grant No. 11805083. C. Y.
Zhang is supported by Natural Science Foundation of China under Grant
No. 11947067.

\end{document}